# A Multiple Data Source Framework for the Identification of Activities of Daily Living Based on Mobile Devices

Ivan Miguel Pires[1], Nuno M. Garcia[1], Nuno Pombo[1], Francisco Flórez-Revuelta[2], Maria Canavarro Teixeira[3,4], Eftim Zdravevski[5] and Susanna Spinsante[6]

[1]Instituto de Telecomunicações, Universidade da Beira Interior, Covilhã, Portugal
[2]Department of Computer Technology, Universidad de Alicante, Spain
[3] UTC de Recursos Naturais e Desenvolvimento Sustentável, Polytechnique Institute of Castelo Branco, Castelo Branco, Portugal
[4] CERNAS - Research Centre for Natural Resources, Environment and Society, Polytechnique Institute of Castelo Branco, Castelo Branco, Portugal
[5] Faculty of Computer Science and Engineering, University Ss Cyril and Methodius, Skopje, Macedonia
[6] Department of Information Engineering, Marche Polytechnic University, Ancona, Italy

```
impires@it.ubi.pt, ngarcia@di.ubi.pt, ngpombo@di.ubi.pt,
     francisco.florez@ua.es, ccanavarro@ipcb.pt,
  eftim.zdravevski@finki.ukim.mk, s.spinsante@univpm.it
```

**Abstract**

Most mobile devices include motion, magnetic, acoustic, and location sensors. They allow the implementation of a framework for the recognition of Activities of Daily Living (ADL) and its environments, composed by the acquisition, processing, fusion, and classification of data. This study compares different implementations of artificial neural networks, concluding that the obtained results were 85.89% and 100% for the recognition of standard ADL. Additionally, for the identification of standing activities with Deep Neural Networks (DNN) respectively, and 86.50% for the identification of the environments with Feedforward Neural Networks. Numerical results illustrate that the proposed framework can achieve robust performance from the data fusion of off-the-shelf mobile devices.

## 1. Introduction

A multiple data source framework for the identification of ADL [1], proposed in [2-4], can be implemented using data acquired from the several sensors available in mobile devices, *e.g.*, accelerometer, magnetometer, gyroscope, microphone, and Global Positioning System (GPS) receiver. These sensors allow the capture of several parameters to enable the automatic identification of the activities. This identification of activities and environments are taken into account the sensing of the characteristics of the movement these activities produce and the environmental sound characteristics associated with the data acquisition period [5]. This framework can integrate the development of a personal digital life coach [6], currently under research.

Based on previous work [7, 8] related to the use of sensors' data for the recognition of ADL and environments, this study enhances both the support to the GPS receiver, and the ability to recognize the driving activity, and a significant number of standing events in comparison with the previous works, including sleeping and watching TV. Since data acquired from the GPS receiver allows the recognition of the geographic location of the user, the proposed framework is capable of distinguishing between different ADL, *e.g.,* the running, walking, walking on stairs, standing, driving and sleeping, and different environments, *e.g.* classroom, gym, kitchen, library, street, hall, watching TV, bedroom. The proposed framework includes several modules, such as the data acquisition performed with a mobile application, the data processing, and the data fusion and classification methods. The data processing module consists of data cleaning and feature extraction. The most important achievement of our framework is related to the adoption of all sensors available in the mobile devices for the recognition either of ADL or environments. It may pave the way not only for a ubiquitous personal life coach but also for healthy aging or disease monitoring and support.

The focus of this study consists in the performance assessment of several methods with a different number of sensors. It allows the adaption of the framework to various mobile devices currently available in the market, as the number of sensors and their capabilities may be different by each smartphone.

This paper continues with the literature review, presented in Section 2, focused on the location sensors for the recognition of ADL and environments. Next, Section 3 presents the methodology used in this study. The results, showed in Section 4, were obtained with the fusion of the location sensors for the recognition of standing activities. Section 5 discusses the results obtained in this study. Finally, Section 6 presents the conclusions of this study.

## 2. Related Work

To date, the recognition of ADL using some sensors available in the mobile devices [9-14] and several classification methods have been widely studied. We can conclude that Artificial Neural Networks (ANN) is one of the most used implementations with the best accuracy [15, 16].

Currently, there are no studies related to the use of the fusion of the data acquired from all sensors available on mobile devices, including accelerometer, gyroscope, magnetometer, microphone, and GPS receiver, for the recognition of ADL and environments [1], but there are some studies using subsets of these sensors. This literature review has its main focus on the use of the GPS receiver for the recognition of ADL and environments where the analysis is available in previous studies [7, 8]. It makes use of the motion and magnetic sensors for the identification of standard ADL, the microphone for the recognition of environmental sounds, and the recognition of standing activities based on the environment

The authors of [17] implemented several methods, such as Support Vector Machine (SVM), Naïve Bayes, ANN, *i.e.,* Multilayer Perceptron (MLP), Logistic Regression, k-Nearest Neighbor (k-NN), Decision Trees, and Rule-Based Classifiers, for the recognition of walking, standing, running, sitting, and going upstairs and downstairs, using the data acquired from accelerometer, magnetometer, GPS receiver, and gyroscope. They extracted the mean and standard deviation related to the accelerometer, gyroscope and magnetometer sensors, and the distance, location, and speed from the GPS receiver, reporting accuracies between 69% and 89% [17].

The data acquired from the accelerometer, GPS receiver and gyroscope for the recognition of ADL, such as standing, walking, running, walking on stairs, and laying activities, were presented in [18]. The authors used several features, including mean, energy, standard deviation, the correlation between axis, and entropy extracted from the motion sensors, and distance, location, and speed obtained from the GPS receiver. That study reported accuracies of 99% with MLP, 96% with logistic regression, 94.2% with a J48 decision tree, and 93.3% with SVM.

In [19], with the data acquired from the accelerometer, the gyroscope, barometer, and the GPS receiver, the authors recognized sitting, standing, washing dishes, walking on stairs, cycling and running with the SVM method. This method was implemented with several features, including mean, standard deviation, and mean squared extracted from the accelerometer and gyroscope sensors, pressure derived from the barometer, and altitude difference in meters and speed from the GPS receiver. Finally, the authors reported an accuracy of 90%.

The authors of [20] used several types of sensors, such as acoustic, location, motion, and medical sensors for the recognition of preparing food, sleeping, standing, jogging, eating, working, and traveling activities. They implemented the Naïve Bayes, C4.5 decision tree, RIPPER, SVM, Random Forest, Bagging, AdaBoost, and Vote methods [20]. The inputs for these methods were the features extracted from the several sensors [20]. The sound features corresponded to the Mel-Frequency Cepstral Coefficients (MFCC), the averages of the spectral centroids, the zero crossing rates, and the Linear Predictive Coding (LPC) values [20]. The distance between to access points is a feature extracted from the Wi-Fi receiver used by the authors of [20]. The features extracted from the GPS receiver were the GPS location identifier, the velocity, and the category of the nearest place [20]. The acceleration features extracted were the elementary activity and energy expenditure obtained by an algorithm [20]. Finally, the Heart-Rate and Respiration-Rate features were minimum, maximum, and average [20]. The reported results obtained by the several methods implemented were 68% for the Naïve Bayes, 66% for the C4.5 decision tree, 72% for the RIPPER, 72% for the SVM, 71% for the Random Forest, 69% for the Bagging, 66% for the AdaBoost, and 77% for the Vote [20].

In [21], with the accelerometer, GPS receiver, camera, and timer used for the recognition of several ADL, including sitting, standing, lying, riding an elevator, walking, dining, going upstairs and downstairs, moving a kettle, washing dishes, preparing a meal, drying hands, moving plates, washing hands, brushing teeth and combing hair, the authors implemented a decision tree as a classification method. The features used were mean and range of Y-axis of the accelerometer, standard deviation of each axis of the accelerometer, sum of intervals of the accelerometer, Signal Magnitude Area (SMA) of amount of variations of the accelerometer, difference of ranges of the accelerometer, interval between X- and Z-axis of the accelerometer, distance, location, and speed, reporting an accuracy between 88.24% and 100% [21].

The SVM was implemented with several features as input, such as the minimum, maximum, mean, standard deviation, correlation between axis and median crossing extracted from the accelerometer data, and the distance, location, and speed obtained from the GPS receiver, to recognize the walking, standing still and running activities, reporting an accuracy around 97.51% [22].

The accelerometer and the GPS receiver were used to recognize standing, traveling by car, traveling by train and walking activities with a J48 decision tree, Random Forest, ZeroR, Logistic, decision table, Radial Basis Function Network (RBFN), ANN, Naïve Bayes, and Bayesian Network [23]. The input features of the methods were the average speed, average accuracy, average rail line closeness, average acceleration, average heading change, magnitudes of the frequency domain, and the signal variance [23]. The average reported accuracies were 85.2% with a J48 decision tree, 85.1% with Random Forest, 84.8% with ZeroR, 84.7% with logistic, 84.6% with decision table, 84.4% with RBFN, 84.4% with MLP, 84.2% with Naïve Bayes, and 84.1% with Bayesian Network [23].

The authors of [24] implemented several methods, including J48 decision tree, MLP, and Likelihood Ratio (LR) for the recognition of going downstairs, jogging, sitting, standing, going upstairs, and walking activities using the accelerometer and GPS receiver. The input features for the methods implemented were the maximum, minimum, mean, standard deviation and zero-crossing rate for each axis for the accelerometer, the correlation between the axis of the accelerometer, and the distance, location, and speed acquired from the GPS receiver [24]. The reported accuracies were 92.4% with a J48 decision tree, 91.7% with MLP, and 84.3% with LR [24].

In [25], the accelerometer and GPS receiver are used for the recognition of standing, driving, walking, running, going upstairs, going downstairs, riding an elevator, and cycling, using Bayesian networks. The input features used are mean, variance, spectral energy, and spectral entropy from the accelerometer, and the location retrieved from the GPS receiver, reporting an accuracy of 95% [25].

Other authors implemented methods for the recognition of ADL using the GPS receiver and other sensors available in off-the-shelf mobile devices. The authors of [26] tested the use of SVM and HMM-based on the data acquired from the accelerometer, gyroscope and GPS receiver, recognizing, standing, walking, running and sitting activities with a reported reliable accuracy.

In [27], traveling by car or train, and cycling activities were recognized based on the International Road Index (IRI) and angle of slope measured by the data acquired from the accelerometer, gyroscope and GPS receiver with reliable accuracy.

The authors of [28] combined the accelerometer with the GPS receiver for the recognition of lying, sitting, standing, and fall activities. The used the following features: Signal Magnitude Area (SMA), Signal Magnitude Vector (SMV) and Tilt Angle (TA) from the accelerometer data, and the distance, location, and speed retrieved from the GPS receiver.

In [29], the GPS receiver, Wi-Fi Positioning System (WPS), GSM Positioning System (GSMPS), and accelerometer were used for the recognition of standing, sitting, lying, walking, jogging, cycling, traveling by bus, train, t taxi or car. These authors applied several features, including the numbers of peaks, the number of troughs, the sum of peaks and troughs, the difference between the maximum peak and the maximum trough, the difference between the maximum and the minimum either peak or trough. The proposed method revealed its effectiveness in terms of energy efficiency (less 53% of battery energy spent than others).

The authors of [30] used only the GPS receiver for the recognition of working, attending lectures, shopping, swimming, training in a gym, playing team sports, visiting friends, eating, going to a pub, to the cinema, to a concert, to the theatre, to a church, and visiting a doctor, based on the density and time-based methods available in the OpenStreetMap (OSM) platform.. Based on different levels of threshold, the accuracies reported by the authors were between 72.2% and 95.4% with the density-based method, and between 66.1% and 69.6% with time-based approach [30].

Following the analysis of the studies available in the literature, Table 1 shows the ADL recognized with GPS receiver and other sensors, sorted in descending order of their respective number of research works. As shown, the standing, walking, sitting, running, going upstairs, going downstairs, and driving/traveling are the most recognized ADL (highlighted in blue background).

*Table 1 - Distribution of the ADL extracted in the studies analyzed.*

| ADL: | Number of Studies: |
|---|---|
| standing | 11 |
| driving/traveling (i.e., car, train, bus, taxi) | 9 |
| walking | 8 |
| sitting | 7 |
| running; going upstairs; going downstairs | 6 |
| lying; cycling | 4 |
| jogging | 3 |
| washing dishes; preparing food; eating; working; riding an elevator | 2 |
| sleeping; dining; moving a kettle; drying hands; moving plates; washing hands; brushing teeth; combing hair; falling; attending lectures; shopping; swimming; training in a gym; playing team sports; visiting friends; going to a pub; going to the cinema; going to a concert; going to the theatre; visiting a doctor; going to church | 1 |

*Table 2 - Distribution of the features extracted in the studies analyzed.*

| Features: | Number of Studies: |
|---|---|
| location | 9 |
| speed | 8 |
| distance | 6 |
| altitude difference in meters; velocity; category of the nearest place; International Road Index (IRI); angle of the slope; Points of Interest (POI) | 1 |

*Table 3 - Distribution of the classification methods used in the studies analyzed.*

| Methods: | Number of Studies: | Average of Reported Accuracy: |
|---|---|---|
| Multilayer Perceptron (MLP) | 4 | 93.53% |
| Logistic Regression | 3 | 93.23% |
| Support Vector Machine (SVM) | 6 | 90.36% |
| Bayesian Network | 2 | 89.55% |
| k-Nearest Neighbor (k-NN) | 1 | 89.00% |
| Rule-Based Classifiers | 1 | 89.00% |
| Decision trees (*i.e.,* J48, C4.5) | 6 | 88.88% |
| ZeroR | 1 | 84.80% |
| Decision Table | 1 | 84.60% |
| Radial Basis Function Network (RBFN) | 1 | 84.40% |
| Likelihood Ratio (LR) | 1 | 84.30% |
| Naïve Bayes | 3 | 83.73% |
| Random Forest | 2 | 78.05% |
| Vote | 1 | 77.00% |
| RIPPER | 1 | 72.00% |
| Bagging | 1 | 69.00% |
| AdaBoost | 1 | 66.00% |

Table 2 shows the features extracted from the data acquired from the GPS receiver in descending order. As observed, the location, speed, and distance are the most common features. On the one hand, the GPS receiver provides the geographic information of the place of activity. On the other hand, speed and distance enable to measure the intensity of the event (shown in text with a blue background).

Finally, Table 3 highlights the most popular methods for the recognition of ADL in descending order of their reported accuracies, such as MLP, Logistic Regression, and SVM (shown in text with a blue background). Therefore, the method that indicates that the best average accuracy in recognition of ADL is with the MLP method, with an average accuracy equals to 93.53%.

### 3. Methods

The development of a multiple data source framework [2-4] for the recognition of ADL enhances the techniques presented in previous studies [7, 8] with the following modules:
- Data acquisition performed by a mobile application;
- Data processing forked in data cleaning and feature extraction methods; and
- Recognition forked in data fusion and classification methods.

#### 3.1. Data Acquisition

The study [8] presented the acquisition of data from the accelerometer, the magnetometer, the gyroscope, the microphone, and the GPS receiver related to several ADL and environments, which is stored in the ALLab MediaWiki [31] and made publicly available for validation of these conclusions and further research.

Considering that the data acquisition occurs for 5 seconds every 5 minutes, and users are active for 16 hours per day, we estimate that the data collection time is around 16 minutes per day. Thus, the proposed method is feasible either in the most sophisticated or low-cost mobile devices.

The ultimate goal of this study is to allow the recognition of standing activities, such as sleeping, driving, and watching TV to map the users' lifestyles.

#### 3.2. Data Processing

This study uses the accelerometer, gyroscope, and magnetometer data, applying the low pass filter [32] for the reduction of the effects of the environmental noise and invalid data. This study also acquires microphone data, and the data is immediately summarized, therefore not compromising the user's privacy, because it only receives raw data.

Based in both the literature and our previous studies [7, 8], the most relevant features extracted in the sensors mentioned above are: the features mentioned in [8] plus the distance traveled, and the environment recognized.

#### 3.3. A fusion of the Data Acquired from Sensors Available in Off-the-shelf Mobile Devices

The features presented in the previous section for the recognition of standing activities are merged, creating different datasets as depicted in Figure 1.

Thus, based on these datasets, three different experiments were performed: (1) using just the accelerometer, (2) using the accelerometer combined with the magnetometer, and (3) using the accelerometer, magnetometer, and gyroscope.

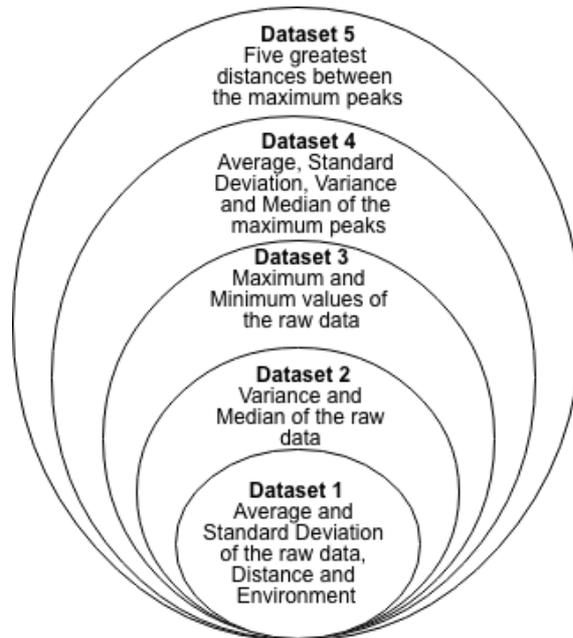

*Figure 1 – Datasets created for the analysis of standing activities, using the distance traveled calculated with the data acquired from the GPS receiver and the previously recognized environment.*

### 3.4. Classification

Based on the literature review presented in section 2, this study focuses on the fusion of the features extracted from the location, motion, mechanical, and acoustic sensors available in the off-the-shelf mobile devices. As observed in the literature, one of the most popular classification methods for the recognition ADL and environments is the ANN, which reports the best accuracy to recognize standing activities.

The implementations used both normalized and non-normalized data with some frameworks and methods implemented and tested in [8], where we tested these methods with different values of maximum training iterations, such as $10^6$, $2\times10^6$ and $4\times10^6$. Figure 2 presents the architecture of the framework for the recognition of ADL and environments.

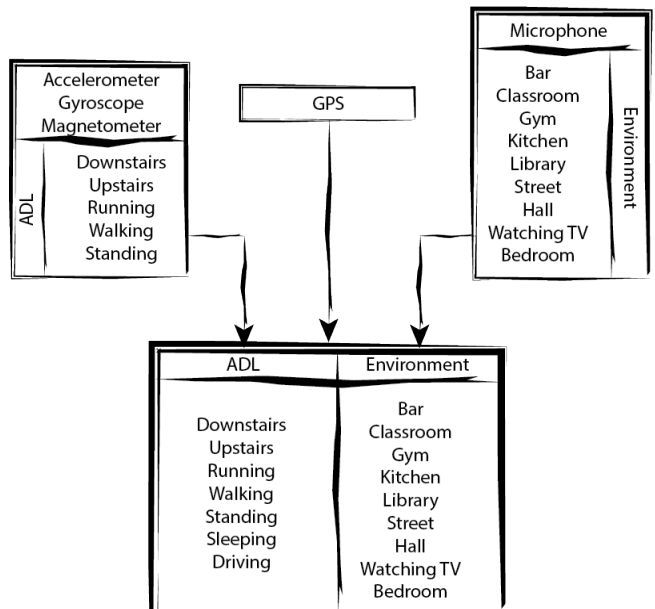

*Figure 2 – ADL and environments recognized by the proposed framework.*

## 4. Results

We benchmark both MLP and FNN with Backpropagation, and the DNN to determine the most accurate method for the recognition of standing activities. The mentioned datasets were created based on 6000 records equally distributed for the proposed standing activities, namely: watching TV, sleeping, and driving.

Based on the analyzed combinations of sensors for the identification of standing activities, the following combinations of sensors' data were defined:
- 1) Environment recognized and the accelerometer and GPS receiver data;
- 2) Sensors' data collected in the mix 1 plus the magnetometer data;
- 3) Sensors' data collected in the combination 2 plus the gyroscope data.

For the identification of standing activities based on the features extracted from the combination 1, our experiments revealed that the results obtained with MLP with Backpropagation are always 100%, except in the dataset 1 (66.67%) with non-normalized data. Also, the results obtained with the FNN with Backpropagation are still 100%, except when in both datasets 4 and 5 (96.67%) with non-normalized data. Finally, the results obtained with the DNN are always 100%, except when is used the dataset 1 (33.37%) with non-normalized data.

Therefore, for the recognition of standing activities based on the features extracted from the combination 2, the obtained results on the implementation of the MLP with Backpropagation are always around 100%, except when is used the dataset 1 (66.72%) with non-normalized data. Similarly, the results obtained with the FNN with Backpropagation are always 100%, except when is used the dataset 1 (94.03%) with non-normalized data. At last, the results obtained with the DNN are still 100% with normalized data and 33.42% with non-normalized data.

Finally, for the recognition of standing activities based on the features extracted from the combination 3, the results revealed an accuracy equals to 100% by MLP method, except when is used the dataset 1 (55.35%) with non-normalized data. Congruently, the results obtained with the FNN with Backpropagation are always 100%, except when is used the dataset 1 (96%) with non-normalized data. Finally, the results obtained with the DNN are still 100% with normalized data and 33.4% with non-normalized data.

Comparing the maximum accuracies reported by the different implementations, Table 4 presents the best results reported with the various combinations of sensors and datasets.

*Table 4 - Best accuracies obtained in recognition of standing activities.*

| | Framework | Dataset (Combination) | Iterations needed for training | Best accuracy achieved (%) |
|---|---|---|---|---|
| **Non-normalized data** | Neuroph | 2 (1) | $10^6$ | 99.97 |
| | | 2 (3) | $10^6$ | 100.00 |
| | | 4 (2) | $10^6$ | 100.00 |
| | Encog | 1 (1) | $10^6$ | 100.00 |
| | | 3 (2) | $10^6$ | 99.97 |
| | | 2 (3) | $10^6$ | 99.98 |
| | Deep learning | 2 (1) | $10^6$ | 100.00 |
| | | 2 (2) | $10^6$ | 33.42 |
| | | 1 (3) | $10^6$ | 33.40 |
| **Normalized data** | Neuroph | 2 (1) | $10^6$ | 100.00 |
| | | 5 (2) | $10^6$ | 100.00 |
| | | 1 (3) | $10^6$ | 100.00 |
| | Encog | 1 (1, 2 or 3) | $10^6$ | 100.00 |
| | Deep learning | 1 (1, 2 or 3) | $10^6$ | 100.00 |

## 5. Discussion

This paper complements the research about the development of a framework for the recognition of ADL and environments [2-4] using all sensors available in the off-the-shelf mobile devices, providing the identification of some ADL with movement, *i.e.,* running, standing, going upstairs, going downstairs, and walking, some user's environments, *i.e.,* bar, classroom, gym, kitchen, library, street, hall, watching TV, and bedroom, and some standing activities, *i.e.,* watching TV, driving and sleeping. Table VIII summarizes the literature on the recognition of ADL and environments. Also, Figure 10 presents the results obtained with the proposed framework for the identification of ADL and environmental sounds using the motion, magnetic, acoustic, and location sensors. The highlighted values in Table V represents the ADL/Environments recognized by the proposed framework.

As shown in Table 5, the results obtained with either the accelerometer or accelerometer combined with magnetometer and gyroscope, revealed the ability of our framework to recognize 5 of 6 ADL presented in the literature (83%). On the other hand, the proposed framework can identify 4 of 8 events presented in the literature (50%) with the accelerometer, magnetometer, gyroscope, and microphone. Finally, 6 of 7 ADL presented in the literature (86%) are recognized by the proposed framework with the accelerometer, magnetometer, gyroscope, microphone, and GPS receiver.

Table 5 - Most recognized ADL and environments based on the literature review distributed by the sensors used (# represents the number of studies available in the literature that identifies the ADL/environment).

| Accelerometer | | Accelerometer Magnetometer Gyroscope | | Accelerometer Magnetometer Gyroscope Microphone | | Accelerometer Magnetometer Gyroscope Microphone GPS receiver | |
|---|---|---|---|---|---|---|---|
| ADL | # | ADL | # | ADL Environment | # | ADL | # |
| Walking | 63 | Walking | 21 | Emergency vehicles | 6 | Resting Standing | 11 |
| Resting Standing | 48 | Going downstairs | 17 | Sleeping | 5 | Driving Travelling | 9 |
| Going upstairs | 45 | Going upstairs | 17 | Walking | 5 | Walking | 8 |
| Going downstairs | 44 | Resting Standing | 16 | Resting Standing | 5 | Sitting | 7 |
| Running | 31 | Running | 13 | Street traffic | 5 | Running | 6 |
| Sitting | 30 | Sitting | 11 | Ocean | 5 | Going upstairs | 6 |
| | | | | Driving | 4 | Going downstairs | 6 |
| | | | | River | 4 | | |

As shown in Table 6, he best accuracies in recognition of standard ADL is 89.51%, in the identification of environments is 86.50%, and, in the identification of standing activities is 100%. Thus, we recommend the implementation of DNN method with normalized data and the application of $L_2$ regularization for the recognition of the standard ADL and standing activities and the implementation of FNN method with non-normalized data for the identification of environments. Also, the average accuracy of the framework for all devices with different combinations of sensors is 91.27%.

Table 6 - Summarization of the accuracy of the final framework for the recognition of ADL and environments.

| Stages of the framework | Accelerometer Microphone GPS | Accelerometer Magnetometer Microphone GPS | Accelerometer Magnetometer Gyroscope Microphone GPS | Average accuracy |
|---|---|---|---|---|
| Recognition of common ADL | 85.89% | 86.49% | 89.51% | 87.30% |
| Recognition of environments | 86.50% | 86.50% | 86.50% | 86.50% |
| Recognition of standing activities | 100.00% | 100.00% | 100.00% | 100.00% |
| Average accuracy | 90.80% | 91.00% | 92% | 91.27% |

## 6. Conclusions

The hardware of off-the-shelf mobile devices includes several sensors that can handle the recognition of ADL and environments in a framework. It combines the data acquired from an enlarged set of sensors available in the mobile devices to develop a framework that adapts their functionalities with the number of sensors available in the equipment used. This paper finished the definition of the methods for the different stages of the framework. The framework starts with the data acquisition, data cleaning and feature extraction methods, and, at the first stage of the recognition of ADL, the framework uses the DNN for the identification of walking, running, standing, going downstairs, and going upstairs. In the second stage, the framework recognizes some environments with the FNN with Backpropagation, and these are a bar, classroom, gym, kitchen, library, street, hall, watching TV, and bedroom. Finally, in the third stage, the framework uses DNN for the recognition of standing activities, and these are watching TV, sleeping, and driving.

The recognition of the ADL and environments is based on the features extracted from the different sensors' data (excluding the microphone and the GPS receiver), such as the five greatest distances between the maximum peaks, the average, standard deviation, variance and median of the maximum peaks, and the standard deviation, average, maximum value, minimum value, variance and median of the raw signal. The features extracted from the microphone data are the 26 MFCC coefficients, the standard deviation, average, maximum value, minimum value, variance, and median of the raw signal. Also, the unique feature obtained from the GPS receiver; the distance traveled, also enables to identify the users' location.

For the development of a framework for the recognition of ADL and environments, we compared three different implementations of ANN, and these are MLP and FNN with Backpropagation and DNN. Our study revealed that an average accuracy of 87.50% in recognition of standard ADL, 86.50% in identification of environments, and 100% in recognition of standing activities. Finally, the average accuracy of the proposed framework is 91.27%. Thus, the proposed framework proves its reliability in the identification of the ADL and its environments.

## Acknowledgments
This work was supported by FCT project **UID/EEA/50008/2013** (*Este trabalho foi suportado pelo projecto FCT UID/EEA/50008/2013*).

The authors would also like to acknowledge the contribution of the COST Action IC1303 – AAPELE – Architectures, Algorithms, and Protocols for Enhanced Living Environments.